\documentclass[final,5p,times,twocolumn]{elsarticle}

\usepackage{graphicx}
\usepackage{amsmath}
\usepackage{amssymb}
\usepackage{xcolor}
\usepackage{soul}
\usepackage{float}
\usepackage{tikz}
\usetikzlibrary{arrows,matrix,positioning}

\usepackage[nottoc]{tocbibind}

\usepackage{hyperref}
\hypersetup{
    colorlinks=true, 
    linktoc=all,     
    citecolor=blue,
    filecolor=black,
    linkcolor=blue,
    urlcolor=blue
}

\biboptions{comma,numbers,compress}

\journal{Computer Physics Communications}

\DeclareUnicodeCharacter{2060}{}
\DeclareUnicodeCharacter{202F}{}
\DeclareUnicodeCharacter{2009}{}
\DeclareUnicodeCharacter{03C0}{}
\DeclareUnicodeCharacter{2215}{}

\begin{document}

\begin{frontmatter}



\title{PLQ$-$Sim: A Computational Tool for Simulating Photoluminescence Quenching Dynamics in Organic Donor$/$Acceptor Blends}


\author[inst1]{Leandro Benatto}
\author[inst1]{Omar Mesquita}
\author[inst2]{Lucimara S. Roman}
\author[inst1,inst3]{Rodrigo B. Capaz}
\author[inst1]{Grazi\^{a}ni Candiotto}
\author[inst2]{Marlus Koehler}

\affiliation[inst1]{organization={Institute of Physics, Federal University of Rio de Janeiro},
            city={Rio de Janeiro},
            postcode={21941$-$4909}, 
            state={RJ},
            country={Brazil}}
            
\affiliation[inst2]{organization={Department of Physics, Federal University of Paraná},
            city={Curitiba},
            postcode={81531-980}, 
            state={PR},
            country={Brazil}}
            
\affiliation[inst3]{organization={Brazilian Nanotechnology National Laboratory (LNNano), Brazilian Center for Research in Energy and Materials (CNPEM)},
            city={Campinas},
            postcode={13083$-$100}, 
            state={SP},
            country={Brazil}}

\begin{abstract}
Photoluminescence Quenching Simulator (PLQ$-$Sim) is a user$-$friendly software to study the photoexcited state dynamics at the interface between two organic semiconductors forming a blend: an electron donor (D), and an electron acceptor (A). Its main function is to provide substantial information on the photophysical processes relevant to organic photovoltaic and photothermal devices, such as charge transfer state formation and subsequent free charge generation or exciton recombination. From input parameters provided by the user, the program calculates the transfer rates of the D$/$A blend and employs a kinetic model that provides the photoluminescence quenching efficiency for initial excitation in the donor or acceptor. When calculating the rates, the user can choose to use disorder parameters to better describe the system. In addition, the program was developed to address energy transfer phenomena that are commonly present in organic blends. The time evolution of state populations is also calculated providing relevant information for the user. In this article, we present the theory behind the kinetic model, along with suggestions for methods to obtain the input parameters. A detailed demonstration of the program, its applicability, and an analysis of the outputs are also presented. PLQ$-$Sim is license free software that can be run via dedicated webserver \href{https://nanocalc.org/}{nanocalc.org} or downloading the program executables (for \textit{Unix}, \textit{Windows}, and \textit{macOS}) from the \href{https://github.com/NanoCalc/PLQ-Sim}{PLQ-Sim} repository on \textit{GitHub}.
\end{abstract}




\end{frontmatter}


\section{Introduction}\label{sec:introduction}

Over the years, the charge transfer (CT) process between electron donor (D) and electron acceptor (A) materials has been widely studied by photoluminescence (PL) measurements \cite{arkhipov2004,arkhipov1995,candiotto2017}.⁠ The PL effect corresponds to the light emission from the radiative exciton recombination after photon absorption in the material. Depending on the material combination, PL can be quenched due to interfacial charge transfer processes \cite{candiotto2017,ohkita2008}⁠. One of the simplest ways to study this effect is by comparing the integrated PL intensity of pristine donor (PL$[$D$]$) with the corresponding quantity for the donor$/$acceptor blend (PL$[$D$/$A$]$) \cite{theander2000,vandewal2012}⁠. In order to efficiently excite the singlet excitons in the donor, an excitation wavelength close to the main absorption band of the donor must be chosen. In order to selectively study the D$\rightarrow$A charge transfer, it is important to minimize the overlap between the donor and acceptor absorption spectrum \cite{benatto2023}. This condition ensures that the majority of the excitons generated after photon absorption are created in the donor phase of the D$/$A blend. The PL quenching efficiency of the donor excitonic luminescence ($PLQ_{D}$) is then

\begin{equation}\label{eq:eq_1}
PLQ_{D}(\%) = 1 - \dfrac{PL\left[D/A\right]}{PL\left[D\right]}.
\end{equation}

\noindent If the excitation wavelength corresponds to acceptor’s main absorption band \cite{cha2018,gerhard2015},⁠ the quenching efficiency of the A phase ($PLQ_{A}$) is quantified by comparing the PL of pristine acceptor ($PL\left[A\right]$) with $PL\left[D/A\right]$

\begin{equation}\label{eq:eq_2}
PLQ_{A}(\%) = 1 - \dfrac{PL\left[D/A\right]}{PL\left[A\right]}.
\end{equation}

\begin{figure}
    \centering
    \includegraphics[width=\linewidth]{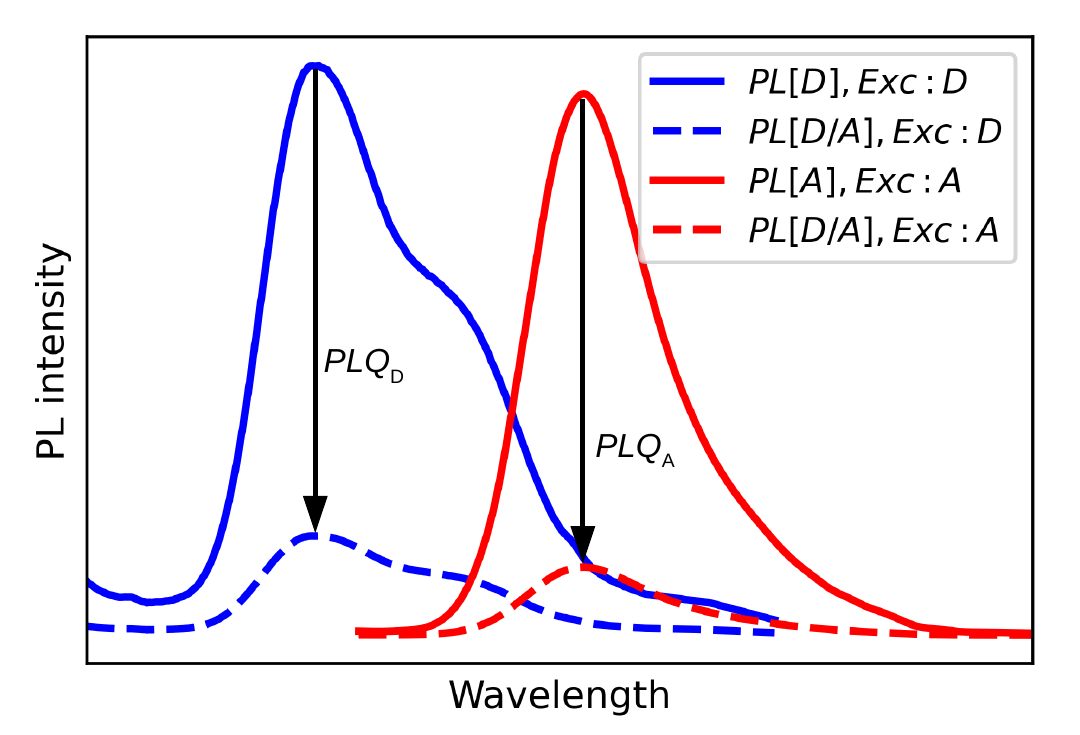}
    \caption{Illustration of measurement of PL intensity versus wavelength after selective excitation of the donor (D) or acceptor (A) materials and blend. The arrows indicate the PL quenching.}
    \label{fig:FIG-1}
\end{figure}

\noindent Figure \ref{fig:FIG-1} illustrates PL measurements using a fluorimeter after selective excitation of the donor or acceptor. The magnitudes of $PL\left[D\right]$, $PL\left[A\right]$ and $PL\left[D/A\right]$ must be obtained by integrating over the entire measured spectrum. $PLQ_{D}$ and $PLQ_{A}$ are determined by the contributions of exciton dissociation or non$-$radiative (NR) exciton recombination at the D$/$A interface. Both processes are intermediated by the creation of a charge transfer (CT) state formed after the electron transfer from the D to A or the hole transfer from A to D. For the $PLQ_{D(A)}$ measurement to capture these effects well, the majority of the generated excitons must reach the D$/$A interface. In bulk heterojunction systems where the D and A materials are mixed, achieving this goal is relatively straightforward. It is essential to control the size of the materials domains to ensure that a significant proportion of excitons encounter the D$/$A interface. In this context, the domain size should be roughly equivalent to the exciton diffusion length. Conversely, in bilayer heterojunctions, where the donor and acceptor are in separate layers, ensuring that most excitons reach the D$/$A interface requires careful consideration of layer thickness. To maximize the chances of excitons reaching the D$/$A interface, the thickness of the layers should be on the order of the exciton diffusion length. In summary, domain size control is pivotal in bulk heterojunctions, while layer thickness control plays a crucial role in bilayer heterojunctions. These considerations are fundamental to understand and optimize PLQ measurements.

The measurement of $PLQ_{D\left(A\right)}$ is a powerful tool to investigate photoluminescence quenching in D$/$A blends. However, the application of this experimental method does not allow to study the details of the processes that contribute to the quenching. Deeper knowledge of the excited state dynamics at D$/$A interfaces by kinetic modeling is fundamental to tailor the properties of the blend aiming at different applications \cite{vezie2019,zhong2020,gerhard2017,sandberg2021,azzouzi2022reconciling,landi2021multiple}⁠. For example, for organic solar cells, exciton dissociation must be maximized and exciton recombination minimized to improve the efficiency of the photovoltaic process. On the other hand, for photo$-$thermal therapy, the maximization of NR exciton recombination is desirable to improve the heat generation produced after photon absorption \cite{grodniski2023}⁠. Therefore, it is important to adopt a general approach that combines the analysis of the experimental data with simulations of $PLQ_{D}$ and $PLQ_{A}$ obtained from a kinetic model of exciton dissociation at the D$/$A interface. The use of this method might reveal the full dynamics of excited states in D$/$A blends which is fundamental to improving the performance of these materials towards a specific use. 

Our group developed a theory to calculate $PLQ_{D}$ and $PLQ_{A}$ based on a kinetic model that considers all possible processes involved in exciton dynamics at the D/A interface \cite{benatto2020,benatto2021-1,benatto2021-2}.⁠ We believe this model can be helpful for research groups interested in the dynamics of charge generation or exciton recombination induced by light excitation of D$/$A blends. To provide broad access to this capability, we developed a free$-$license and easy to handle software, Photoluminescence Quenching Simulator (PLQ-Sim). We also extend the original model to consider the time evolution of state populations following a similar procedure recently presented in ref. \cite{landi2021multiple}. To perform a simulation, basic information about the materials forming the system (such as energy levels, dielectric constant, reorganization energy, and singlet exciton lifetime, for instance) must be provided as input parameters. If some information is missing, default values for the class of material studied can be used. After running the simulation, the rates of the physical processes involved in the charge dynamics at the D$/$A interface (like the charge transfer or charge recombination rates, for instance) are provided as output. This information is crucial to characterize the photophysical process as a whole and to propose ideas for system optimizations. Therefore, our program aims to help the experimentalist to fine$-$tune fabrication process to maximize the PL quenching. PLQ$-$Sim is a free license program compatible with \textit{Windows}, \textit{Unix}, and \textit{macOS} operating systems developed in Python programming language.

\section{Theoretical Methods}\label{sec:meth}
\subsection{Kinetics of photoluminescence quenching}\label{sec:KPQ}

\begin{figure}
    \centering
    \includegraphics[width=\linewidth]{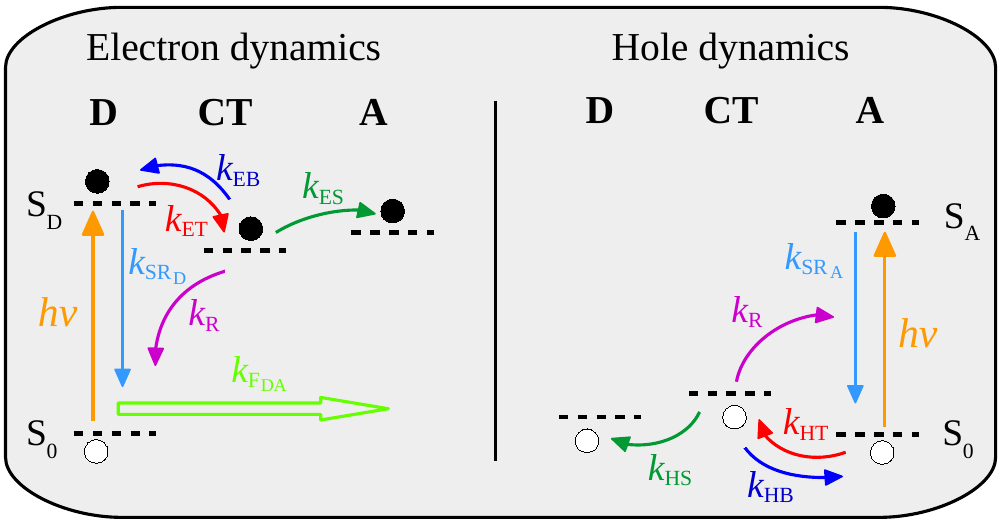}
    \caption{Simplified charge dynamics diagram after light absorption in the donor (left) or acceptor (right) material.}
    \label{fig:FIG-2}
\end{figure}

The model describes, using a simple one$-$electron picture, the electron and hole kinetics at the D$/$A interface. In addition to charge transfer, it also takes into account Fluorescence (or F\"{o}rster) resonance energy transfer (FRET) \cite{benatto2023,candiotto2020}.  After donor excitation, four processes might occur with their corresponding characteristic rates: electron transfer ($k_{ET}$), singlet exciton recombination ($k_{SR_{D}}$), donor$-$donor$-$FRET ($k_{F_{DD}}$) or donor$-$acceptor$-$FRET ($k_{F_{DA}}$). If electron transfer takes place, a charge transfer state (CT) is formed. From this intermediary excitation, three new kinetic steps might follow: electron separation ($k_{ES}$), electron return to regenerate the initial singlet state ($k_{EB}$) or decay to the ground state by charge recombination from CT ($k_{R}$). If instead of electron transfer the donor$-$acceptor$-$FRET takes place as a first step, the dynamics of the hole needs to be taken into account with analogous processes as those described above for electrons. The presence of donor$-$acceptor$-$FRET thus links the $PLQ_{D}$ with the hole kinetics at the acceptor. Based on Ref. \cite{zhu2014}⁠, Figure \ref{fig:FIG-2} illustrates the charge and energy transfer dynamics that will be studied here. The main objective of our model is to quantify the steady state values of $PLQ_{D}$ and $PLQ_{A}$ using the transfer rates of the characteristic processes as input parameters.

We start the description of the model assuming that there is on averaged $N$ excited donor molecules near an acceptor. In an interval $dt$, $n_{1}$ of those $N$ donors can transfer an electron to the acceptor to form a CT state (group 1). Alternatively, besides the electron transfer, $n_{2}$ of those donors have the additional possibility of performing an energy transfer by FRET to a nearby acceptor (group 2). From the above assumptions $n_{1} + n_{2} = N$. Thus, for group 1 molecules, the time derivative of singlet exciton concentration, $P_{S_{D}}$, is

\begin{equation}\label{eq:eq_3}
\dfrac{dP_{S_{D}}}{dt} = I - \left(k_{SR_{D}}+k_{ET}\right)P_{S_{D}} + k_{EB}P_{CT},
\end{equation}

\noindent where, $P_{CT}$ is the concentration of CT states and  $I$ is the exciton generation rate in the donor at the D$/$A interface. This rate involves the excitons produced by direct photon absorption or resulting from the net flux of excitations transferred by FRET from (to) other donors. Using the above assumptions, time evolution of singlet state concentration for the group 2 of donors is given by

\begin{equation}\label{eq:eq_4}
\dfrac{dP_{S_{D}}}{dt} = I - \left(k_{SR_{D}}+k_{ET}\right)P_{S_{D}} + k_{EB}P_{CT} -k_{F_{DA}}P_{S_{D}},
\end{equation}

\noindent where the term $k_{F_{DA}}P_{S_{D}}$ was added to Eq. (\ref{eq:eq_3}) from Eq. (\ref{eq:eq_4}). By combining these equations one gets the total time variation of the singlet state concentration in the $N$ molecules

\begin{equation}\label{eq:eq_5}
\dfrac{dP_{S_{D}}}{dt} = -\dfrac{n_{2}}{N}k_{F_{DA}}P_{S_{D}} + c,
\end{equation}

\noindent where $c\equiv I - \left(k_{SR_{D}}+k_{ET}\right)P_{S_{D}} + k_{EB}P_{CT}$. If $N$ is sufficiently high, the probability that one excited donor will belong to group 2 is $p_{F_{DA}}\approx n_{2}/N$. Hence Eq. (\ref{eq:eq_5}) can be written as a function of $p_{F_{DA}}$

\begin{equation}\label{eq:eq_6}
\dfrac{dP_{S_{D}}}{dt} = -p_{F_{DA}}k_{F_{DA}}P_{S_{D}} + c.
\end{equation}

\noindent Upon charge transfer, the time$-$dependent concentration of CT state at the D$/$A heterojunction is related to $P_{S_{D}}$ by

\begin{equation}\label{eq:eq_7}
\dfrac{dP_{CT}}{dt} =k_{ET}P_{S_{D}} -\left(k_{EB}+k_{ES}+k_{R}\right)P_{CT}.
\end{equation}

\noindent In writing Eq. (\ref{eq:eq_7}), we  neglected the formation of CT states by direct absorption of light due to the low oscillation strength of this transition. Within the steady state approximation, Eq. (\ref{eq:eq_6}) and Eq. (\ref{eq:eq_7}) gives

\begin{equation}\label{eq:eq_8}
I = \left(k_{SR_{D}}+k_{ET}\right)P_{S_{D}} - k_{EB}P_{CT}+p_{F_{DA}}k_{F_{DA}}P_{S_{D}}
\end{equation}

\begin{equation}\label{eq:eq_9}
P_{CT}=\dfrac{k_{ET}P_{S_{D}}}{k_{EB}+k_{ES}+k_{R}}.
\end{equation}

\noindent Substituting Eq. (\ref{eq:eq_9}) into Eq. (\ref{eq:eq_8}) one finds

\begin{equation}\label{eq:eq_10}
P_{S_{D}}=f_{D}I,
\end{equation}

\noindent where

\begin{equation}\label{eq:eq_11}
f_{D}=\dfrac{k_{EB}+k_{ES}+k_{R}}{\left(k_{SR_{D}}+k_{ET}+p_{F_{DA}}k_{F_{DA}}\right)\left(k_{EB}+k_{ES}+k_{R}\right)-k_{EB}k_{ET}}.
\end{equation}

After the D$-$A$-$FRET, the exciton at the acceptor can either recombine emitting a photon or be transferred to the CT state. Hence, the population of excitons in the acceptor also influences $PLQ_{D}$ even for selective illumination of donor. Following the same reasoning used to derive Eq. (\ref{eq:eq_10}), one can analogously write the population of (FRET induced) singlet excitons in the acceptor $P_{S_{A},F}$ as

\begin{equation}\label{eq:eq_12}
P_{S_{A},F}=f_{A}k_{F_{DA}}P_{S_{D}}=f_{A}k_{F_{DA}}f_{D}I,
\end{equation}

\noindent where

\begin{equation}\label{eq:eq_13}
f_{A}=\dfrac{k_{HB}+k_{HS}+k_{R}}{\left(k_{SR_{A}}+k_{HT}\right)\left(k_{HB}+k_{HS}+k_{R}\right)-k_{HB}k_{HT}}.
\end{equation}

\noindent In Eq. (\ref{eq:eq_12}), it is assumed that energy transfer from A to D is not allowed, so that $k_{F_{AD}}=0$.

In the absence of acceptors (hence assuming only donor molecules), the time derivative of singlet state concentration $P'_{S_{D}}$ at the same position considered in Eq. (\ref{eq:eq_10}) would give

\begin{equation}\label{eq:eq_14}
\dfrac{dP'_{S_{D}}}{dt}=I'-k_{SR_{D}}P'_{S_{D}},
\end{equation}

\noindent where  I'  is the rate of exciton generation in a donor$-$only material. Again assuming steady-state conditions, Eq. (\ref{eq:eq_14}) gives,

\begin{equation}\label{eq:eq_15}
I'=k_{SR_{D}}P'_{S_{D}},
\end{equation}

\noindent or

\begin{equation}\label{eq:eq_16}
P'_{S_{D}}=\dfrac{I'}{k_{SR_{D}}}.
\end{equation}

\noindent Since the selective illumination of the donor can generate photoluminescence of the acceptor due to FRET, we define the donor’s PL quenching as

\begin{equation}\label{eq:eq_17}
\begin{split}
PLQ_{D}&=1-\left(\dfrac{\left(1-p_{F_{DA}}\right)P_{S_{D}}+p_{F_{DA}}P_{S_{A,F}}}{P'_{S_{D}}}\right),\\
PLQ_{D}&=1-\left(\dfrac{\left(1-p_{F_{DA}}\right)P_{S_{D}}+p_{F_{DA}}f_{A}k_{F_{DA}}P_{S_{D}}}{P'_{S_{D}}}\right),
\end{split}
\end{equation}

\noindent where we use Eq. (\ref{eq:eq_12}). If the acceptor is inefficient to quench the excitons and $I\approx I'$, then $P'_{S_{D}}\approx \left(1-p_{F_{DA}}\right)P_{S_{D}}+p_{F_{DA}}P_{S_{A,F}}$ from energy conservation. Eq. (\ref{eq:eq_17}) then gives $PLQ_{D}\approx 0$. Alternatively, if the acceptor is efficient to dissociate the excitons created in the donor, then $P_{S_{D}}\approx 0$ and, from Eq. (\ref{eq:eq_17}), $PLQ_{D}\approx 1$. Note that the D$-$A$-$FRET is not dominant when $p_{F_{DA}}\approx 0$ so that the exciton quenching is determined mainly by electron transfer from the donor to the acceptor. On the other hand, if $p_{F_{DA}}\approx 1$ the exciton quenching will be determined by a  sequential process that involves the D$-$A$-$FRET and the following transfer of holes to the donor. Using Eqs. (\ref{eq:eq_12}) and (\ref{eq:eq_16}), after some algebra, one gets

\begin{equation}\label{eq:eq_18}
PLQ_{D}=1-k_{SR_{D}}f_{D}\left(\left(1-p_{F_{DA}}\right)+p_{F_{DA}}f_{A}k_{F_{DA}}\right).
\end{equation}

\noindent To completely assess the quenching dynamics described by Eq. (\ref{eq:eq_18}), it is necessary to determine $p_{F_{DA}}$. We will assume that $p_{F_{DA}}$ depends on the process that deactivates (or reactivates) the singlet donor state \cite{chou2015}. Under this assumption $p_{F_{DA}}$ will be given by:

\begin{equation}\label{eq:eq_19}
p_{F_{DA}}=\dfrac{k_{F,DA}}{k_{F,DA}+k_{F,DD}+k_{SR,D}+k_{ET}-k_{EB}}.
\end{equation}

\noindent In the case of $p_{F_{DA}}\approx 0$ the formula for $PLQ_{D}$ in Eq. (\ref{eq:eq_18}) is reduced to the expression:

\begin{equation}\label{eq:eq_20}
PLQ_{D}=1-\dfrac{P_{S_{D}}}{P'_{S_{D}}}=1-k_{SR_{D}}f_{D}.
\end{equation}

\noindent Finally, considering the selective excitation of the acceptor, we can repeat the same reasoning described above to calculate $PLQ_{A}$. Taking into account the absence of FRET from acceptor to the donor ($k_{F_{AD}}$), one can use a similar expression to Eq. (\ref{eq:eq_20}) to write $PLQ_{A}$

\begin{equation}\label{eq:eq_21}
PLQ_{A}=1-\dfrac{P_{S_{A}}}{P'_{S_{A}}}=1-k_{SR_{A}}f_{A}.
\end{equation}

\noindent See that from the kinetic model it is possible to obtain $PLQ_{D}$ and $PLQ_{A}$ using the transfer rates involved in the process.

\subsection{Transfer rates calculation}\label{sec:TRC}

The Marcus$/$Hush equation\cite{coropceanu2007} is used in the program to calculate the characteristic frequencies involved in the charge transfer process

\begin{equation}\label{eq:eq_22}
k=\dfrac{4\pi^{2}}{h}\dfrac{\beta^{2}}{\sqrt{4\pi\lambda k_{B}T}}\exp\left[-\dfrac{\Delta G^{\ddag}}{k_{B}T}\right],
\end{equation}

\noindent where $k_{B}$, $T$, $\lambda$, and $\beta$ are respectively the Boltzmann constant, temperature, reorganization energy, and electronic coupling (transfer integral). The activation energy for charge transfer, $\Delta G^{\ddag}$, is given by \cite{oberhofer2017,merces2021}

\begin{equation}\label{eq:eq_23}
\Delta G^{\ddag} = \dfrac{\left(\lambda + \Delta G\right)^2}{4\lambda}
\end{equation}

\noindent where $\Delta G$ is the Gibbs free energy (driving force) of the charge transfer reaction, approximated here as the energy difference between final and initial states. As it will be demonstrated later, the reorganization energy, electronic coupling and energy levels are input parameters to the program. Specifically, the Marcus$/$Hush equation is used to obtain the frequency of the following rates: $k_{ET}$, $k_{EB}$, $k_{ES}$, $k_{HT}$, $k_{HB}$, $k_{HS}$ and $k_{R}$.

\begin{figure}
    \centering
    \includegraphics[width=\linewidth]{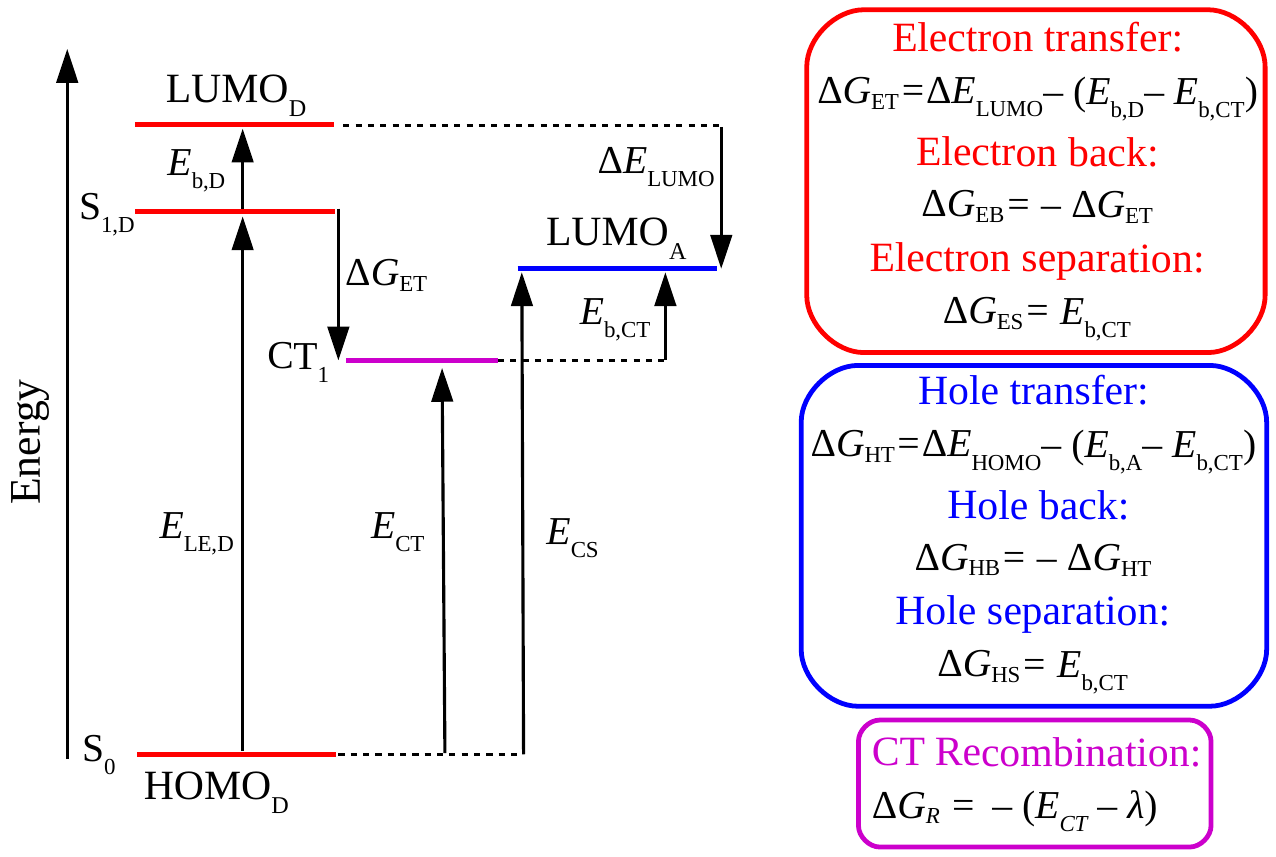}
    \caption{Scheme of energy levels and driving forces for electron and hole dynamics.}
    \label{fig:FIG-3}
\end{figure}

Figure \ref{fig:FIG-3} illustrates the energies involved in the kinetic processes depicted in Figure \ref{fig:FIG-1} as derived from the various energy levels of the system \cite{nakano2019anatomy,qian2018}. The definitions of driving forces involved in the electron and hole dynamics at the D$/$A interface are also shown in Figure \ref{fig:FIG-3}. Note that, contrary to a common misconception that the driving force for electron ($\Delta G_{ET}$) or hole transfer ($\Delta G_{HT}$) is solely given by the energy offset between LUMOs ($\Delta E_{LUMO}$) or HOMOs ($\Delta E_{HOMO}$), it also involves the binding energy of singlet ($E_{b,D}$ and $E_{b,A}$) and CT states ($E_{b,CT}$). Only under the exceptional condition  $E_{b,D}=E_{b,A}=E_{b,CT}$, the driving forces are simplified to $\Delta G_{ET} = \Delta E_{LUMO}$ and $\Delta G_{HT} = \Delta E_{HOMO}$ as assumed in several works. The $\Delta G$ for electron or hole back jump to recreate the singlet state is simply the negative of the respective transfer driving force. Considering the final step of exciton dissociation, the electron or hole separation, the driving force is given by the binding energy of the CT state. From the findings of Chen \textit{et al}. \cite{chen2016}⁠, the $\Delta G$ for CT recombination ($\Delta G_{R}$) can be estimated as the negative of the CT state energy subtracted from the reorganization energy. When $\Delta G_{R}$ is estimated using this procedure, the $k_R$ found using the Marcus$/$Hush theory is similar to the rate derived using the Marcus$-$Levich$–$Jortner model that considers the thermal population of high frequency vibrational modes \cite{benatto2019-2}.

The FRET rates between donors ($k_{F_{DD}}$) and between donor and acceptor ($k_{F_{DA}}$) are  additional program inputs that must be provided by the user. Their quantification can be easily performed by a free program developed by our group called FRET$-$Calc \cite{benatto2023}. If FRET effects are negligible in the D$/$A blend under study, the FRET rates can be set to zero in the input file. Additionally, if the user intends to study only $PLQ_{A}$, there is no need to provide the FRET rates. Finally, the singlet exciton recombination rate in each phase of the blend ($k_{SR_{D}}$ and $k_{SR_{A}}$) is given by the inverse of the singlet exciton lifetime accessed from spectroscopy measurements in pristine materials, for instance.

\subsection{Disorder effects}\label{sec:DE}

To better describe the charge transfer dynamics at the D$/$A interface, the program also consider effects of diagonal disorder (energy levels fluctuations) and off$-$diagonal disorder (electronic coupling fluctuations between adjacent molecules due to random orientations) \cite{coropceanu2007}.⁠ The diagonal disorder is considered by selecting a random value from a Gaussian distribution with mean zero and standard deviation $\sigma$ (defined in the input) that is added to the driving forces. 

For the calculations of off$-$diagonal effects, the user must define the maximum value for electronic coupling, $\beta_{max}$, as an initial input. The program will then use the equation $\beta = \beta_{max}\cos \left(\phi\right)$ in the calculations to determine an specific value of the electronic coupling. The angle $\phi$ will be randomly chosen between $0^{\circ}$ and a user$-$defined maximum value ($\phi_{max}$) in the range $0^{\circ} - 90^{\circ}$. For example, if the user chooses $\phi_{max}=90^{\circ}$, adjacent molecules will be considered randomly between the face$-$on configuration ($\phi_{min}=0^{\circ}$ and $\beta=\beta_{max}$) and edge$-$on configuration ($\phi_{max}=90^{\circ}$ and $\beta=0$).

When disorder is considered, the transfer rates, $PLQ_{D}$ and $PLQ_{A}$ become unique for each realization of parameters. It is then necessary to calculate average quantities to characterize the charge dynamics for a particular D$/$A system. To obtain numerical averages with sufficient precision, it is recommended that the values of $PLQ_{D}$ and $PLQ_{A}$ are averaged over at least $10^{4}$ runs of parameters’ realization (defined in the input) of the simulation \cite{gerhard2017,benatto2021-1}. Following this procedure, the average transfer rates, $PLQ_{D}$ and $PLQ_{A}$ are calculated.
	
\subsection{Time evolution of state populations}\label{sec:TESP}

The formalism described above was developed to quantify the photoluminescence quenching when the sample is submitted to a continuous excitation represented by the rate $I$. However, in many situations a more complete characterization of the excited state dynamics can be obtained from time$-$resolved PL measurements. This motivated us to extend our model in order to broaden the applicability of the program.

Hence, besides the steady state PL quenching, the program can also calculate the time evolution of state population using the frequency rates specified above. In essence, this feature of our code can be applied to study experiments in which the sample is fully excited during a very short interval of time. The excited state decay kinetics is then obtained from the evolution of properties such as the intensity of emission response, for instance.  
	
Modeling of this kind of experiment is done by generalizing the previous model to explicitly consider the evolution of molecule populations in the ground and charge separated states. This generalization is done by writing a set of coupled differential equations that describes the inter$-$dependent charge kinetics in the donor and in the acceptor. Basically these equations are derived from the general expression \cite{cupellini2018}

\begin{equation}\label{eq:eq_24}
\dot{P}_{i}\left(t\right)=\sum_{i}\left(-k_{i\rightarrow j}P_{i}\left(t\right)+k_{j\rightarrow i}P_{j}\left(t\right)\right),
\end{equation}

\noindent where $P_{i(j)}(t)$ is the $i$ $(j)$ state population at time $t$. The appropriated solution of Eq. \ref{eq:eq_24} is found by assuming that only singlet excited states are initially populated by a very short light pulse. Then, it follows that the $P_{i(j)}(t)$ decay obeys a multiexponential kinetics, with various effective rates. For a local acceptor excitation there are four coupled differential equations

\begin{equation}\label{eq:eq_25}
\left\{
\begin{aligned}
\dfrac{dP_{S_{A}}}{dt} &= -\left(k_{SR_{A}}+k_{HT}\right)P_{S_{A}}+k_{HB}P_{CT},\\
\dfrac{dP_{CT}}{dt} &= k_{HT}P_{S_{A}}-\left(k_{HB}+k_{HS}+k_{R}\right)P_{CT},\\
\dfrac{dP_{CS}}{dt} &= k_{HS}P_{CT},\\
\dfrac{dP_{GS}}{dt} &= k_{SR_{A}}P_{S_{A}}+k_{R}P_{CT},
\end{aligned}
\right.
\end{equation}

\noindent where the initial condition is $P_{S_{A}} = 1$. Including the FRET process, six coupled differential equations are then needed to describe the kinetics derived from the donor local excitation

\begin{equation}\label{eq:eq_26}
\left\{
\begin{aligned}
\dfrac{dP_{S_{D}}}{dt} &=-\left(k_{SR_{D}}+k_{ET}+k_{F_{DA}}\right)P_{S_{D}}+k_{EB}P_{CT_{D}},\\
\dfrac{dP_{CT_{D}}}{dt} &=k_{ET}P_{S_{D}}-\left(k_{EB}+k_{ES}+k_{R}\right)P_{CT_{D}},\\
\dfrac{dP_{S_{A}}}{dt} &= k_{F_{DA}}P_{S_{D}}-\left(k_{SR_{A}}+k_{HT}\right)P_{S_{A}}+k_{HB}P_{CT_{A}},\\
\dfrac{dP_{CT_{A}}}{dt} &= k_{HT}P_{S_{A}}-\left(k_{HB}+k_{HS}+k_{R}\right)P_{CT_{A}},\\
\dfrac{dP_{CS}}{dt} &= k_{ES}P_{CT_{D}}+k_{HS}P_{CT_{A}},\\
\dfrac{dP_{GS}}{dt} &=k_{SR_{D}}P_{S_{D}}+k_{SR_{A}}P_{S_{A}}+k_{R}P_{CT_{D}}+k_{R}P_{CT_{A}},\\
\end{aligned}
\right.
\end{equation}

\noindent where the initial condition is $P_{S_{D}}=1$. In Eq. (\ref{eq:eq_26}), $P_{CT_{D}}$ is the population of CT state generated from donor exciton, whereas $P_{CT_{A}}$ is the population of CT state generated from acceptor exciton produced by FRET. The total population of CT state is given by, $P_{CT}=P_{CT_{D}}+P_{CT_{A}}$. The time interval considered for simulation is defined by the user in the program’s input file.

\subsection{Limitations of the model }\label{sec:ML}

Although the model outlined above is efficient for  simulating exciton dynamics at the D$/$A interface, it has some limitations that are pointed out below. 1) The simulations implicitly assume that there are well$-$defined donor and acceptor domains and that all generated excitons reach the D/A interface. There is no need to define the heterojunction type before the simulation. Therefore, it is advisable to compare theoretical outcomes with experimental data wherein the predominant fraction of excitons does indeed reach the D$/$A interface. 2) Photophysical processes occurring far from the D$/$A interface region cannot be studied because the model only simulates the exciton dynamics in this region. 3) The model only considers singlet excitons, assuming that the dynamics of triplet excitons do not affect the PL quenching efficiency. It is meaningless to study with the model systems in which the flow of triplet excitons to the D$/$A interface is relevant. 4) As the simulation approach is based on hopping model, whose intermolecular electronic coupling is much smaller than intramolecular charge reorganisation energy, tunneling effects that can assist the electron transfer are not considered. Therefore, systems with high electronic delocalization will not be well described by the model. In this way, we are assuming that the decoherence factor due to interaction with the vibrational degrees of freedom completely destroys the excitonic coherence, which leads to localization of the exciton. 5) The model was developed to simulate D$/$A systems that are not exposed to an external electric field. Therefore, it is not possible to reproduce $PLQ_{D\left(A\right)}$ measurements under the influence of an external electric field.

\subsection{Methods for obtaining input parameters}\label{sec:MOIP}

In this section we will briefly describe some popular methods found in the literature to obtain the input parameters necessary to run the PLQ$-$Sim program. We emphasize that the methods presented below are just suggestions (it does not have the intention to be a complete summary of available experimental$/$theoretical methods). Due to the extensive literature in the area, it is likely that there are other approach to obtain the same parameters.

\subsubsection{Energy levels}\label{sec:EL}

The ionization potential (IP) and electron affinity (EA) are essential materials properties to calculate the driving forces for electron (hole) transfer. In principle, they can be determined by a combination of ultraviolet photoelectron spectroscopy (UPS) and inverse photoemission spectroscopy (IPES) \cite{karuthedath2021}.⁠ Cyclic voltammetry (CV) is another alternative to access these parameters \cite{yang2019}.⁠ Following Koopmans' theorem, the highest occupied molecular orbital (HOMO) energy can be considered as $-$IP while the lowest unoccupied molecular orbital (LUMO) energy represents $-$EA \cite{bredas2014}.

\subsubsection{Energy of charge transfer state}\label{sec:ECTS}

The energy of charge transfer state ($E_{CT}$) can be obtained from the blend's electroluminescence spectra \cite{list2018,wouk2018}⁠. If this technique is not easily available, the simple approximation $E_{CT}=E_{LUMO_{A}}-E_{HOMO_{D}}$, can be used \cite{benatto2021-2,zong2018,wang2017}.

\subsubsection{Singlet exciton recombination lifetime}\label{sec:SERL}
The singlet exciton recombination lifetime of pristine materials can be accessed using time-resolved photoluminescence (TRPL) spectroscopy \cite{yang2017}.

\subsubsection{Relative dielectric constant}\label{sec:RDC}
The relative dielectric constant ($\varepsilon_{r}$) of each material is also required as input. There are two common ways to obtain $\varepsilon_{r}$ from experiments. The first is by performing capacitance measurements using impedance spectroscopy \cite{li2021,zhang2019}⁠. $\varepsilon_{r}$ is then calculated with the help of the equation $\varepsilon_{r}=Cd/ \varepsilon_{0}A$ \cite{liu2018}, where $C$ is the capacitance, $d$ is the thickness of the film, $\varepsilon_{0}$ is the permittivity of free space, and $A$ is the device area. The second technique involves the application of spectroscopic ellipsometry to find the real and imaginary components of the refractive index. In this case $\varepsilon_{r}=n^{2}+k^{2}$, where $n$ is measured in the longer wavelength limit to minimize absorption \cite{chandran2017}. From density functional theory (DFT) calculations, the Clausius$-$Mossotti equation can used to obtain $\varepsilon_{r}$ \cite{schwenn2011,benatto2019-1}.

\subsubsection{Reorganization energy}\label{sec:RE}

The reorganization energy for charge transfer is generally calculated using the adiabatic potential energy surfaces of the neutral and charged molecules \cite{coropceanu2007,oberhofer2017,hutchison2005}.

\subsubsection{Electronic coupling}\label{sec:EC}

The electronic coupling (transfer integral) between each pair of molecular orbitals involved in the electron and hole transfer can be obtained from simulations by applying the fragment orbital method \cite{oberhofer2017,gorelsky2001}.⁠

\subsubsection{Exciton binding energy of pristine materials}\label{sec:EBEPM}

The energy difference between the fundamental (or transport) band gap ($E_{fund}$) and optical band gap ($E_{opt}$) provides the exciton binding energy: $E_{b}=E_{fund}- E_{opt}$. The fundamental band gap is the energy of free charge carriers (electron/hole): $E_{fund} = IP - EA$.  The optical band gap is the energy of the lowest excited state of the semiconductor and can be extracted from the absorption onset in the UV$-$vis spectrum \cite{zhang2022}.

DFT and time$-$dependent DFT (TD$-$DFT) calculations can also be used to obtain the gas$-$phase exciton binding energy ($E_{b,gas}$). The first step is to apply DFT to obtain the ground$-$state geometry of isolated materials. Then, from single point calculations, the total energies of the cationic ($E_{+}$), anionic ($E_{-}$), and neutral ($E_{0}$) states can be determined. From these energies it is possible to calculate $IP = E_{+} - E_{0}$ and $EA = E_{0} - E_{-}$. The optical gap can be obtained from TD$-$DFT calculations and it corresponds to the energy of the lowest energy electronic transition excited by a single photon absorption. This procedure has been widely used in the literature \cite{benatto2019-2,benatto2019-1,zhu2018,lee2015,benatto2023enhancing}.⁠ The exciton binding energy in solid$-$state can be estimated by  $E_{b,solid} = E_{b,gas}/\varepsilon_{r}$.

\subsubsection{Exciton binding energy of charge transfer state}\label{sec:EBECTS}

It is often assumed that the strong dielectric stabilization of the charge transfer (CT) state compensates the coulomb binding of a Frenkel exciton in the solid state, making $E_{b,A}\approx E_{b,CT}$ \cite{karuthedath2021}.⁠ We found in our previous work that there is a poor agreement between the simulated and measured $PLQ$ using this approximation in the calculations \cite{benatto2021-2}⁠. There was a considerable improvement of the theoretical description if one slightly adjusts the $E_{b,CT}$ in relation to $E_{b,A}$. In fact, this is an interesting way to estimate the value of $E_{b,CT}$. Extensive research has been conducted to investigate the CT exciton binding energy in diverse D$/$A blends through various techniques, revealing substantial variations \cite{gerhard2017,athanasopoulos2019binding,dong2019binding}.

\subsection{Recommendations for users}\label{sec:REC}

We emphasize that due to the number of parameters involved in the model calculations, the theoretical estimates of $PLQ_{D}$ and $PLQ_{A}$ must always be confronted with experimental results. From this comparison, some input parameters, which may vary from material to material, can be fine$-$tuned to produce a better correlation between theory and experiment. Finally, we also emphasize that the theoretical model presented here has already been properly tested for several combinations of donor and acceptor materials showing good correlation with experiments and providing valuable insights about the physics behind exciton dissociation in systems under consideration \cite{benatto2021-2,grodniski2023}.

\section{Software architecture, implementation and requirements}\label{sec:soft-arch}

PLQ$-$Sim is a free-license program. The binary files of this code (for \textit{Windows}, \textit{Unix} and \textit{macOS} operational system) are available for download at \href{https://github.com/NanoCalc/PLQ-Sim}{PLQ-Sim} repository on \textit{GitHub} or can be run \textit{via} dedicated webserver \href{https://nanocalc.org/}{nanocalc.org}. The program is implemented in Python 3 (v. 3.6)\cite{van2009}⁠ and makes use of four Python libraries, namely Pandas \cite{pandas2010}⁠, NumPy \cite{harris2020},⁠ and SciPy \cite{virtanen2020}⁠ for data manipulation and Matplotlib \cite{hunter2007}⁠ for data visualization. The software was designed to be very user$-$friendly and take up little disk space, around 80 MB.

\section{Program and Application}\label{sec:program-app}

\subsection{Input and basic demonstration}\label{sec:input}

\begin{figure}
    \centering
    \includegraphics[width=\linewidth]{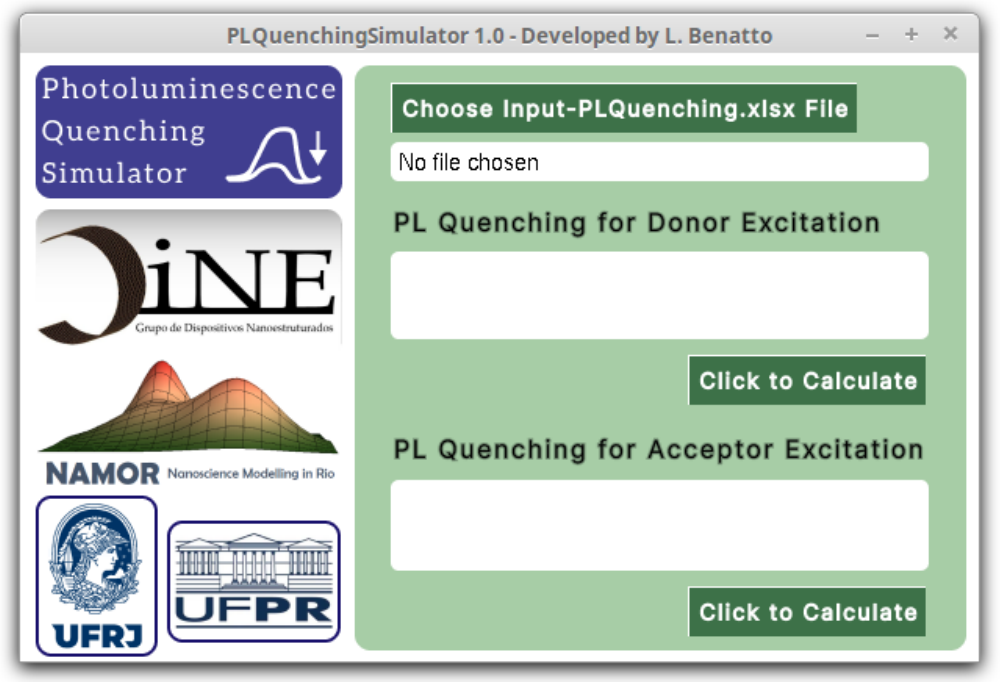}
    \caption{PLQ$-$Sim graphic interface.}
    \label{fig:FIG-4}
\end{figure}

In this section, we will provide a practical example of how to perform a calculation using PLQ$-$Sim. When the program starts, a graphical user interface is loaded. In order to utilize the interface, an input file must be provided as shown in Figure \ref{fig:FIG-4}. The input file contains information on the system that will be simulated and must be in the \textit{xlsx} (Excel file) format.

\begin{figure}[!t]
    \centering
    \includegraphics[width=\linewidth]{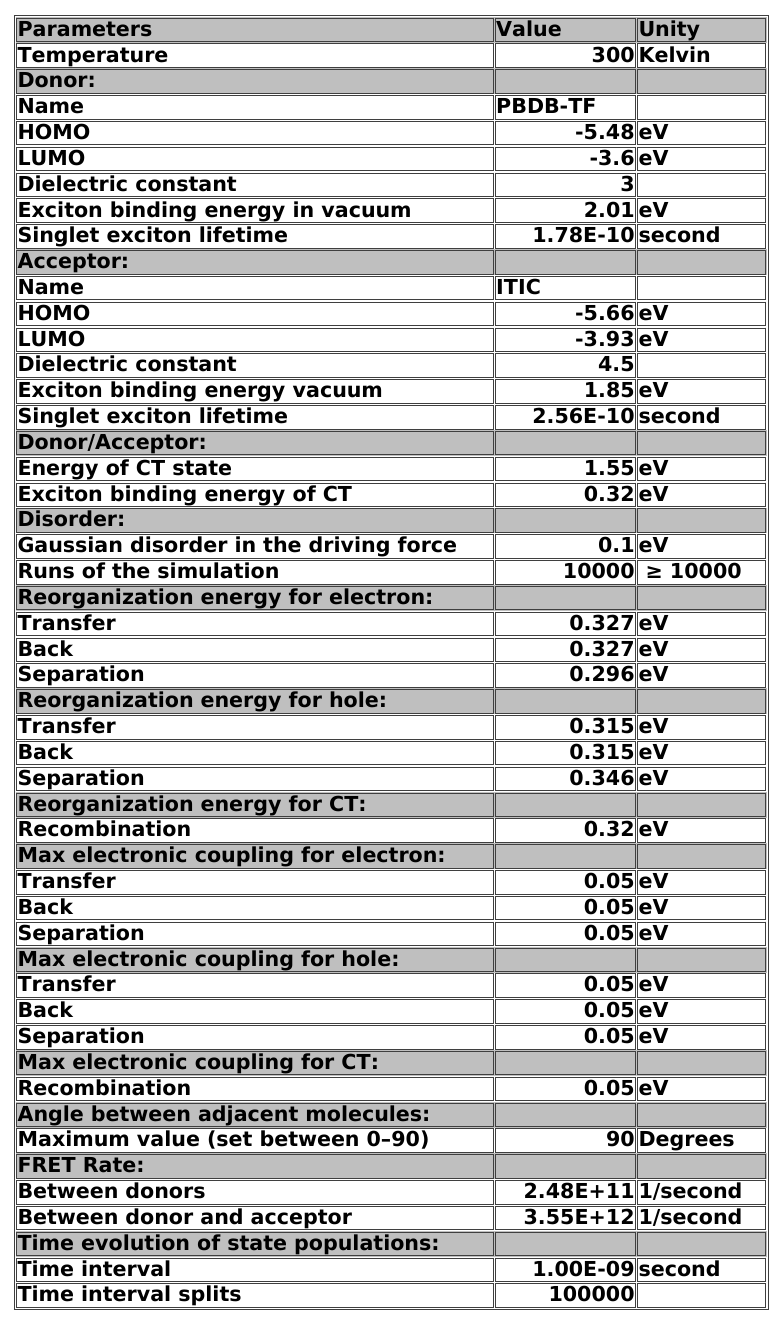}
    \caption{Input$-$PLQ$-$PBDB$-$TF$-$ITIC.xlsx file with the input parameters. }
    \label{fig:FIG-5}
\end{figure}

\begin{figure}[!t]
    \centering
    \includegraphics[width=\linewidth]{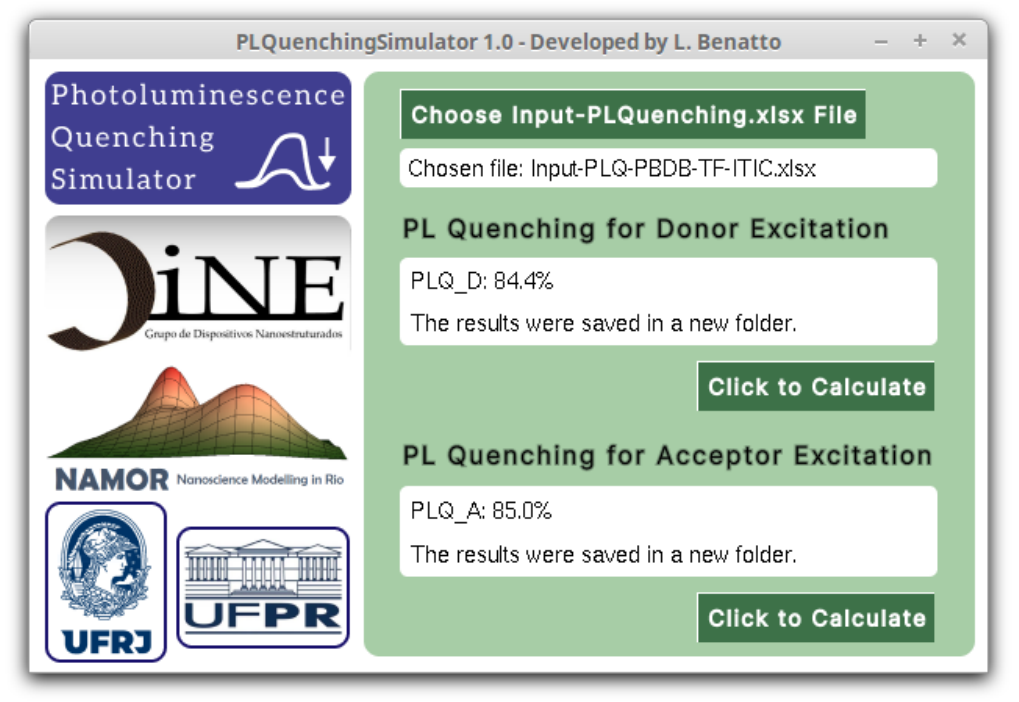}
    \caption{Provided the input file, the program calculates $PLQ_{D}$ and $PLQ_{A}$. Additionally, other informations of the simulated systems are printed in the files Results\textunderscore Donor.dat and Results\textunderscore Acceptor.dat}
    \label{fig:FIG-6}
\end{figure}

A modifiable input file will be included with the program as an illustrative example (see Figure \ref{fig:FIG-5}). The default parameters utilized in this file were derived from literature and describe a D$/$A blend of PBDB$-$TF⁠ donor polymer \cite{zhang2015} (also known as PM6 and PBDB$-$T$-$2F \cite{zheng2020}⁠) and the non$-$fullerene acceptor (NFA) ITIC⁠ molecule \cite{zhao2016}. Those materials are current used in organic solar cells. The molecular energy levels of the materials considered here were obtained using CV measurements \cite{yang2019}⁠. The singlet exciton recombination lifetimes of 178 ps for PBDB$-$TF \cite{karuthedath2021}⁠ and 256 ps for ITIC \cite{yang2017}⁠ were applied to find the singlet exciton recombination rates $k_{SR_{D}}$  and $k_{SR_{A}}$. Dielectric constant, gas$-$phase exciton binding energy, CT state energy and reorganization energy were obtained from Ref.\cite{benatto2021-2}. For diagonal disorder, $\sigma=$ 0.1 eV was considered, which is a standard value for organic semiconductors \cite{gerhard2015,rubel2008,chen2021}. The maximum value for electronic coupling was set to 50 meV for all processes, which is a typical value for the face$-$on configuration of adjacent molecules \cite{chen2021,do2016,aldrich2019,firdaus2020}. The donor$-$acceptor FRET rate were calculated using the FRET$-$Calc software \cite{benatto2023}. Another input information is $E_{b,CT}$. The specific value used in the input file was found by adjustment between theory and experiment. Finally, a total number of $10^{4}$ simulation runs was set as default in the input file. This number was found to give sufficiently accurate results (deviation in $PLQ$ is less than 1\% between simulations using the same set of input data). For the PBDB$-$TF$/$ITIC blend,  $PLQ_{D}$=82\% and $PLQ_{A}$=85\% were experimentally determined \cite{yang2019}⁠ while $PLQ_{D}$=84\% and $PLQ_{A}=$85\% were calculated using the PLQ$-$Sim code (see Figure \ref{fig:FIG-6}) with the set of parameters in Figure \ref{fig:FIG-5}. As the theoretical $PLQ_{D}$ and $PLQ_{A}$ are consistent with the experiments, it is possible to use the additional software outputs to analyze the details of the charge dynamics on the D$/$A interface.

\subsection{Complementary outputs to PLQ$_{D}$ and PLQ$_{A}$}\label{sec:outputs}

\begin{figure}[!htb]
    \centering
    \includegraphics[width=\linewidth]{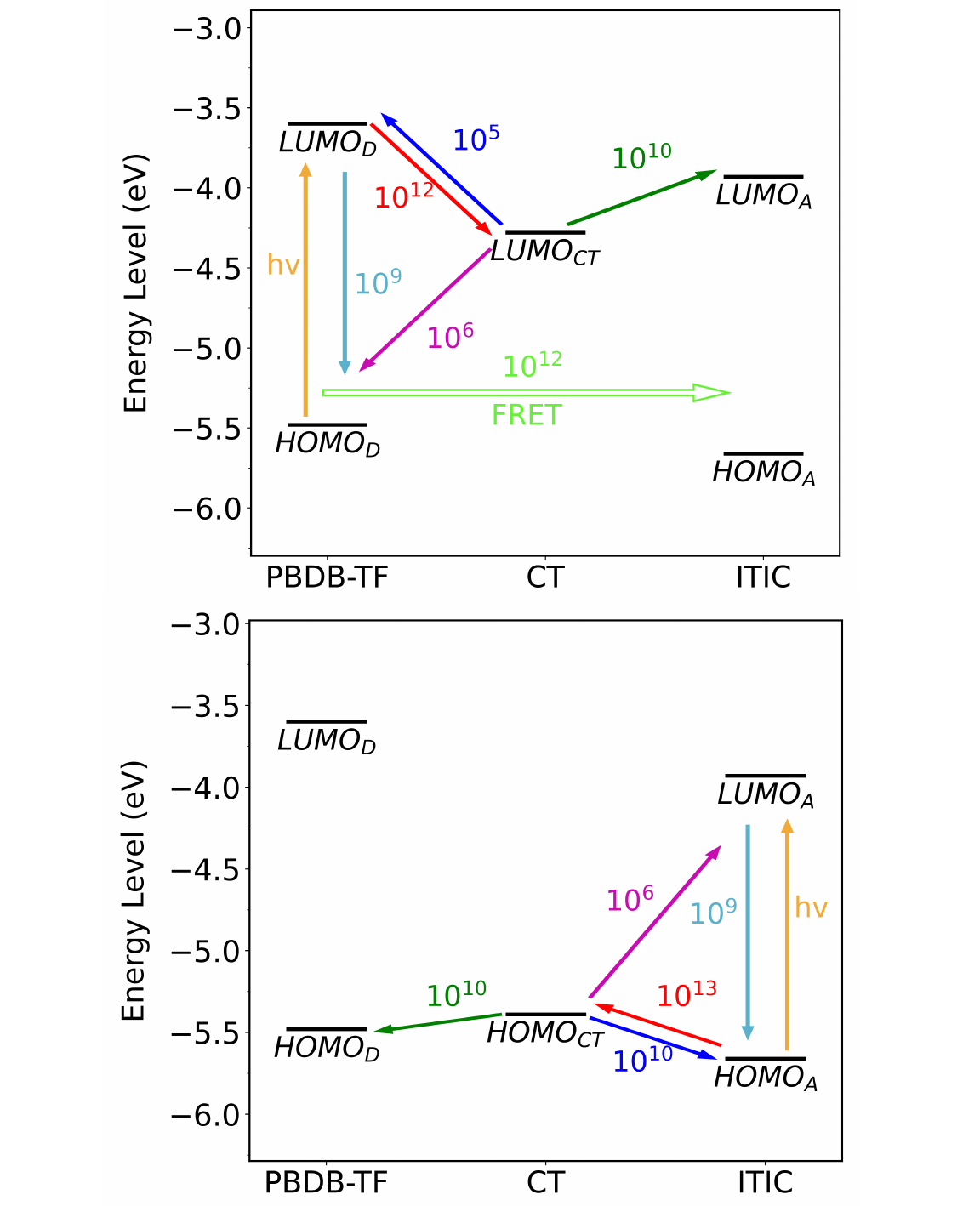}
    \caption{Energy$-$level representation and averaged rates for the electron dynamics (top) and for the hole dynamics (bottom).}
    \label{fig:FIG-7}
\end{figure}

 Relevant data are printed in the Results\textunderscore Donor.dat and Results\textunderscore Acceptor.dat files. These data, such as transfer rates and quenching efficiency, are obtained by averaging over all specific information that are explicitly listed in the Results\textunderscore Donor\textunderscore Steps.dat and Results\textunderscore Acceptor\textunderscore Steps.dat files for each simulation step. In addition, two graphs that illustrate the electron (hole) dynamics at the D$/$A interface (with the respective magnitudes of transfer rates) are automatically generated, see Figure \ref{fig:FIG-7}.

The energy levels diagram and the schematic kinetics among excitons, CT states, and free charge carriers depicted in Figure \ref{fig:FIG-7} are very useful since it can provide the general scenario that determines the exciton dissociation or recombination. For example, in the Figure \ref{fig:FIG-7} (top), $k_{ET}$ and $k_{F_{DA}}$ have the same order of magnitude (10$^{12}$ s$^{-1}$) showing a clear competition between the electron transfer and the FRET phenomena to produce PL quenching. Furthermore, both rates are three orders of magnitude higher than  $k_{SR_{D}}\approx$ 10$^{9}$ s$^{-1}$, which is an important feature leading to $PLQ_{D}$ enhancement. Another interesting result to be highlighted is that the electron separation rate $k_{ES}\approx$ 10$^{10}$ s$^{-1}$ is four orders of magnitude higher than the CT non$-$radiative recombination rate $k_{R}\approx$ 10$^{6}$ s$^{-1}$ and five orders of magnitude higher than the electron back rate to the singlet state $k_{EB}\approx$ 10$^{5}$ s$^{-1}$. The magnitudes of these rates indicate that the donor excitation can create a significant density of CT and CS states at the D$/$A interface.

In addition, $k_{HS}$ and $k_{HB}$ in Figure \ref{fig:FIG-7} (bottom) have the same order of magnitude (10$^{10}$ s$^{-1}$), which suggests that there is a strong competition between the hole separation and back transfer processes. It is also possible to verify that $k_{HT}\approx$ 10$^{13}$ s$^{-1}$ is one order of magnitude greater than $k_{ET}\approx$ 10$^{12}$ s$^{-1}$. In other words, the D$/$A interface is more efficient to transfer holes to the donor than electrons to the acceptor. Another important insight derived from Figure \ref{fig:FIG-7} is that the singlet recombination rate for acceptor excitons, $k_{SR_{A}}\approx$ 10$^{9}$ s$^{-1}$, is three orders of magnitude greater than the non$-$radiative recombination rate of the CT state, $k_{R}\approx$ 10$^{6}$ s$^{-1}$. Since the hole transfer is very efficient, a low $k_{R}$ is essential to enhance the photovoltaic effect. On the other hand, considering the application of D$/$A blends for photothermal conversion, faster exciton recombination and high heat generation might be found if $k_{R} > k_{SR_{D(A)}}$ \cite{grodniski2023}.

\subsection{Outputs for time evolution of state populations}\label{sec:outputsTE}

\begin{figure}[!t]
    \centering
    \includegraphics[width=\linewidth]{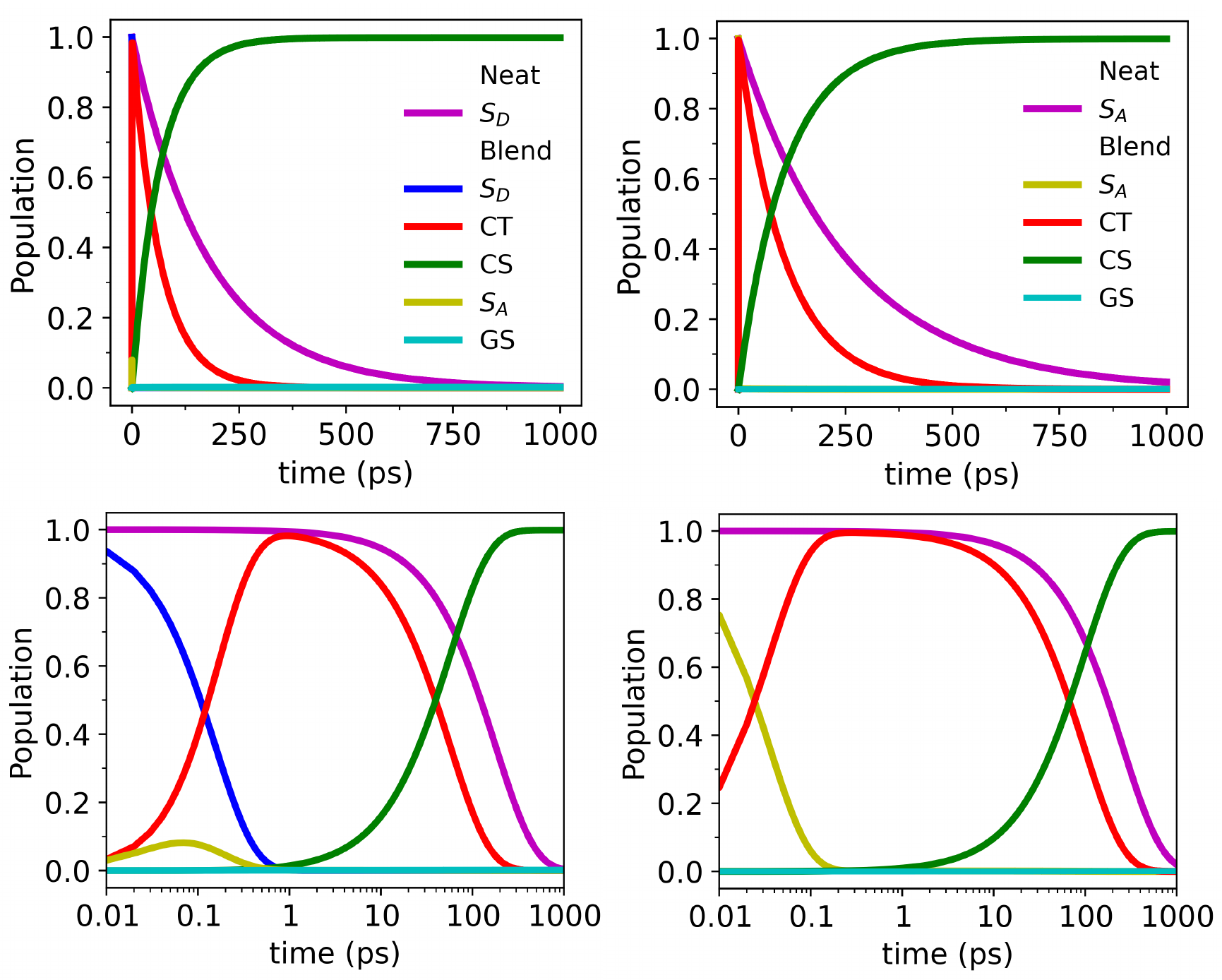}
    \caption{Time evolution of state populations in linear scale (top) and logarithmic scale (bottom). Results on the left correspond to initial excitation in the donor while results on the right correspond to initial excitation in the acceptor.}
    \label{fig:FIG-8}
\end{figure}

An additional output shows the time evolution of each state population. This tool is exemplified in Figure \ref{fig:FIG-8} for linear scale (top) and for logarithmic scale (bottom). Results on the left corresponds to donor excitation while those on the right to acceptor excitation. For comparison, we also calculate the population evolution of the single state that would be present in a pure donor or acceptor  (\textit{i. e.}, disregarding the terms related to CT and CS in Eq. \ref{eq:eq_25} and \ref{eq:eq_26}). The data showed in the Figure 8 are saved by the program in the files Time\textunderscore Evolution\textunderscore Donor.dat and Time\textunderscore Evolution\textunderscore Acceptor.dat. Note in Figure \ref{fig:FIG-8} that the blend population of $S_{D}$ and $S_{A}$ has a very sudden drop to zero (in less than 1 ps), a behavior that is in clear contrast to the kinetics observed for pure (neat) materials.  This steep decay in the blend is due to the fast charge transfer rate ($k_{ET}$ and $k_{HT}$) from the singlet state to form the CT state. In case of donor excitation, the high FRET rate induces the growth of the $S_A$ population which is rapidly extinguished by the fast hole transfer ($k_{HT}$). The population of the CT state is converted into the CS state over time because $k_{ES}$ and $k_{HS}$ are much faster than $k_{R}$. As a consequence, the population of the GS state remains close to zero during the whole time interval analyzed since the majority of the excitons initially created are converted into free charges in this period. The simulated dynamics of the excited states are relevant to complement experimental studies of transient absorption (TA) on D$/$A blends \cite{zhong2020,classen2020}.

\begin{figure}[!t]
    \centering
    \includegraphics[width=\linewidth]{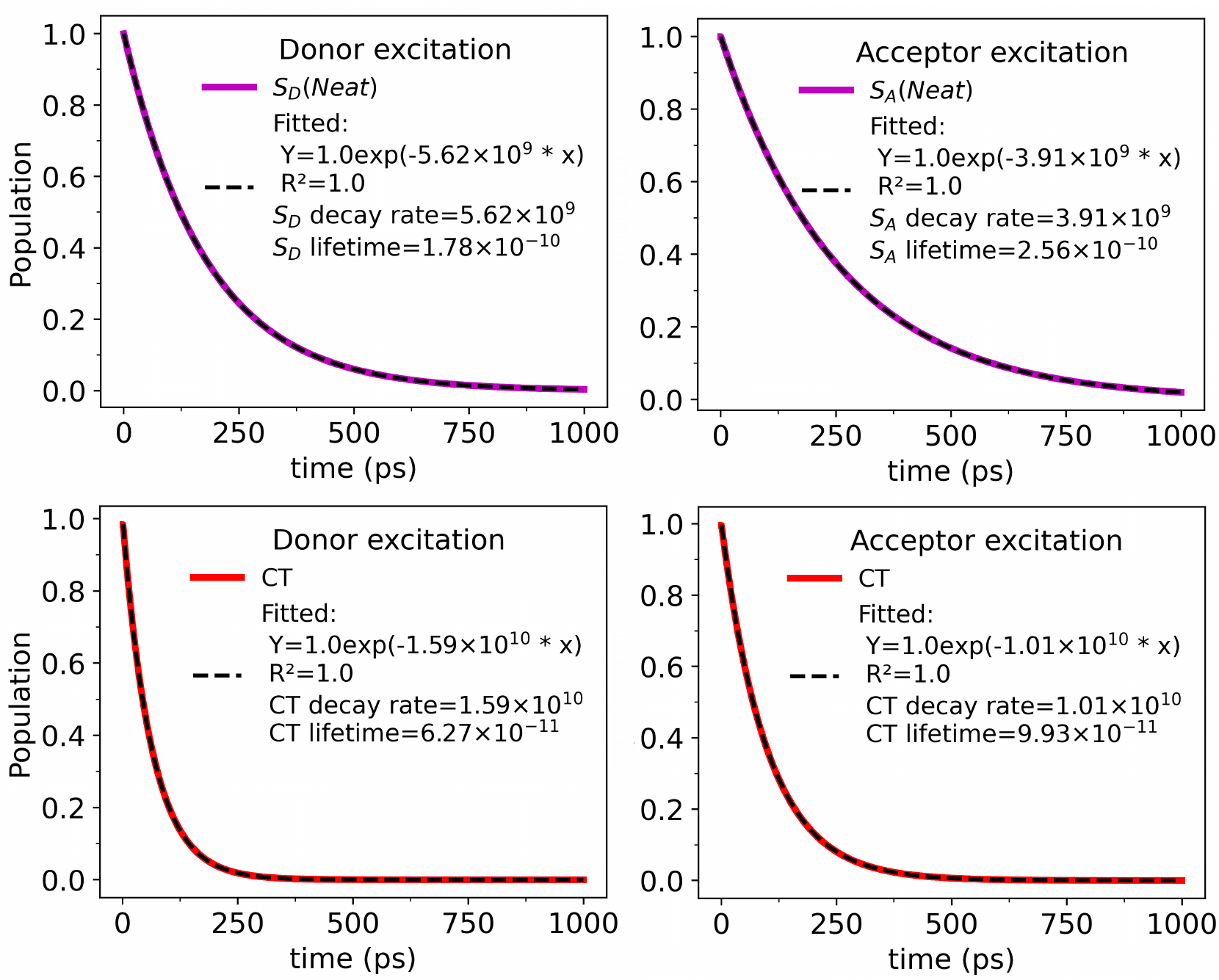}
    \caption{Monoexponential fit of the $S_{D}$(Neat), $S_{A}$(Neat) and CT populations over time. Results on the left correspond to initial excitation in the donor while results on the right correspond to initial excitation in the acceptor.}
    \label{fig:FIG-9}
\end{figure}

A monoexponential fit of the time evolution displayed in Figure \ref{fig:FIG-8} enables the determination of an effective characteristic exciton lifetime and decay rate of the CT state. The program performs this analysis automatically and the results are shown in Figure \ref{fig:FIG-9}. The exciton lifetime and decay rate for singlet states in the neat materials (S$_{D}$(Neat) and S$_{A}$(Neat)) are also calculated for comparison. As expected, the exciton lifetime of S$_{D(A)}$(Neat) (top left and top right) is the same as the user$-$defined one set in the input file (obviously the decay rate $k_{SR_{D(A)}}$ is the inverse of this time). More interesting is the analysis of the CT state decay. It is described in the bottom left (right) of Figure \ref{fig:FIG-9} for donor (acceptor) excitation. Note that the decay rate of the CT state is close to the dissociation rate for the CS state, $k_{E(H)S}$ for donor (acceptor) excitation. Again, this feature of the blend originates from  the slow CT decay ($k_{R}$) to ground state (GS). It is important to mention that the CT state lifetime calculated with the model implemented in the PLQ$-$Sim code is consistent with recent experimental studies involving different polymer$/$NFAs blends \cite{dong2021}.

\subsection{Exploring the software features}\label{sec:features}

\begin{figure}[!t]
    \centering
    \includegraphics[width=\linewidth]{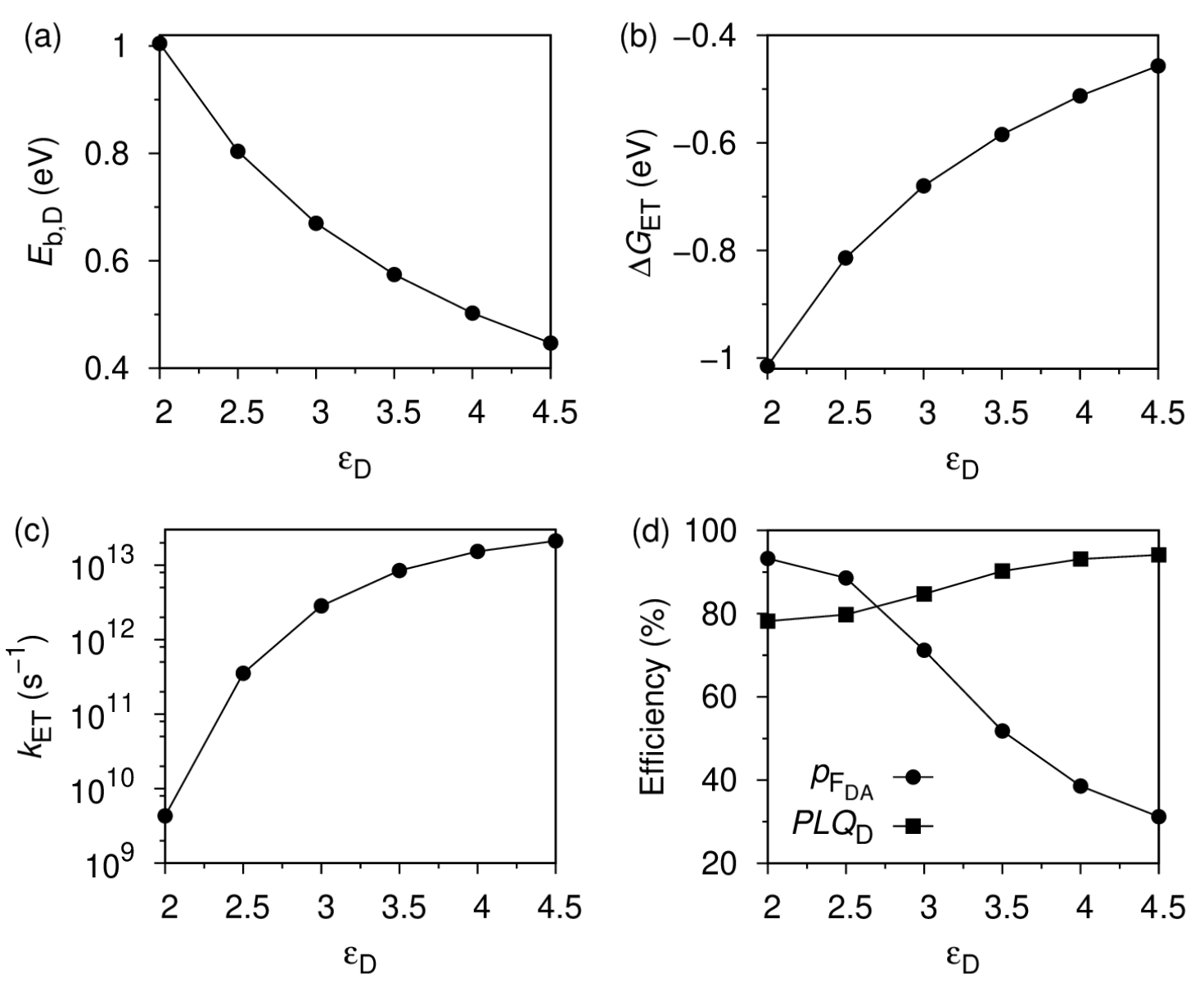}
    \caption{Influence of the dielectric constant of donor ($\varepsilon_{D}$) on the (a) exciton binding energy of donor, (b) driving force for electron transfer, (c) electron transfer rate in log scale and (d) FRET probability and PL quenching efficiency of the donor excitonic luminescence.}
    \label{fig:FIG-10}
\end{figure}

An interesting way to inspect the program is by varying the input parameters and checking their effect on the calculated quantities. Below we change some selected input parameters as examples of exploratory strategies to test the software’s capabilities. Let's begin by varying the relative dielectric constant of donor ($\varepsilon_{D}$). One consequence of increasing $\varepsilon_{D}$ is the reduction of the exciton binding energy, a feature that is evident in  Figure \ref{fig:FIG-10}a. The increase of $\varepsilon_{D}$ also rises the driving force for electron transfer ($\vert \Delta G_{ET}\vert$, Figure \ref{fig:FIG-10}b that can approach the reorganization energy for electron transfer. Here it is important to point out that the blend is in the Marcus inverted region (MIR) for electron transfer, characterized by $\vert \Delta G_{ET}\vert > \lambda_{ET}$ in the framework of semiclassical Marcus theory. Therefore, the effect of $\vert \Delta G_{ET}\vert \rightarrow \lambda_{ET}$ decreases the activation energy for electron transfer which maximizes $k_{ET}$, as can be seen in Figure \ref{fig:FIG-10}c. One can also observe in Figure \ref{fig:FIG-10}d that the FRET probability from donor to acceptor is high in the interval where $k_{ET}$ is low. As $k_{ET}$ becomes higher, the FRET probability is greatly reduced, evidencing the negative correlation between charge transfer and energy transfer. Finally, higher values of $\varepsilon_{D}$ in Figure \ref{fig:FIG-10}d tend to maximize $PLQ_{D}$, mainly by the enhancing $k_{ET}$. Similar conclusions relative to the process of hole transfer and $PLQ_{A}$ can be derived by considering variations of the acceptor relative dielectric constant ($\varepsilon_{A}$).

\begin{figure}[!t]
    \centering
    \includegraphics[width=\linewidth]{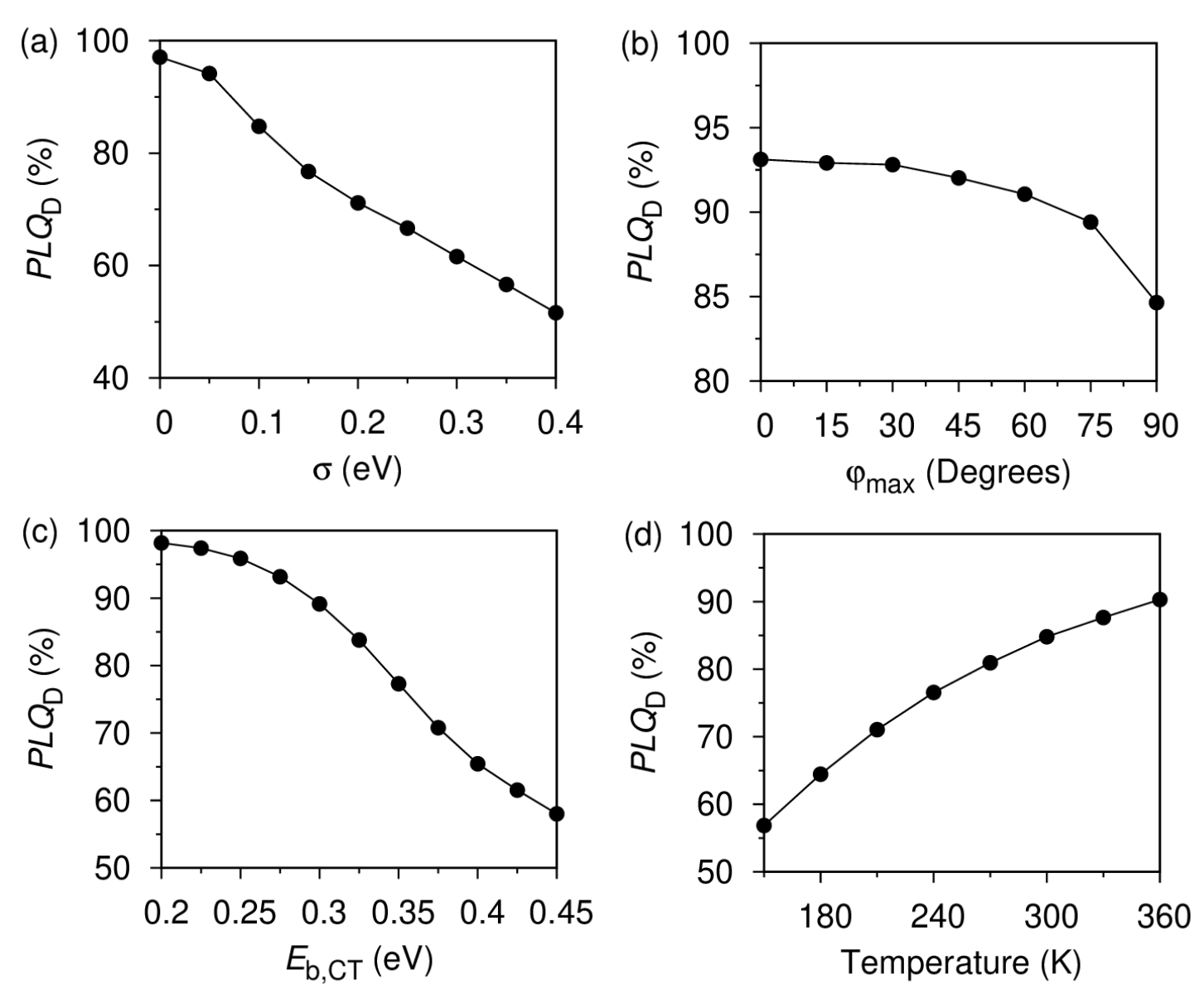}
    \caption{Behavior of $PLQ_{D}$ versus (a) Gaussian disorder in the driving forces, (b) maximum angle between adjacent molecules, (c) binding energy of CT state and (d) temperature.}
    \label{fig:FIG-11}
\end{figure}

Discussion about the role played by disorder to determine some key properties of organic semiconductors has been the subject of a long debate in the literature \cite{candiotto2017,coropceanu2007,rubel2008,peumans2004,koehler2006,liu2020,bakulin2016}⁠. The PLQ$-$Sim code can help to answer these questions by opening the possibility to study, for instance, how $PLQ_{D(A)}$ depends on the disorder. This kind of analysis is exemplified in Figure \ref{fig:FIG-11}a. It is clear that fluctuations in the driving forces associated to disorder tends to decrease $PLQ_{D}$. 

The influence of the maximum angle between adjacent molecules ($\phi_{max}$) on $PLQ_{D}$ is shown in Figure \ref{fig:FIG-11}b. Note that $PLQ_{D}$ is maximum for $\phi_{max}=0^{\circ}$ but slowly decreases until approximately $\phi_{max}=30^{\circ}$. For higher angles, there is a more pronounced decrease until $\phi_{max}=90^{\circ}$. Therefore, the increase in both kinds of disorder (diagonal, Figure \ref{fig:FIG-11}a, and off$-$diagonal, Figure \ref{fig:FIG-11}b) is detrimental to $PLQ_{D}$ for the blend under consideration.

Another parameter that has a great impact on $PLQ_{D}$ is the binding energy of the CT state. In Figure \ref{fig:FIG-11}c one can see that the $E_{b,CT}$ increase causes a regular drop of $PLQ_{D}$  mainly originated by a continuous $k_{ES}$ reduction. When $k_{ES}$ becomes much smaller than the charge transfer rate (10$^{12}$ s$^{-1}$), exciton recombination is favored. Considering temperature effects in organic systems, there are reports in the literature of its beneficial impact on the exciton dissociation process \cite{gerhard2015,bakulin2016,ma2021}.⁠ In Figure \ref{fig:FIG-11}d, $PLQ_{D}$ indeed increases with temperature in agreement with the literature.

\section{Conclusions}
In summary, the theoretical background, a detailed demonstration, the applicability, and usefulness of the PLQ$-$Sim program were described in detail. The program can be a powerful auxiliary tool to study the photoexcited state dynamics in D$/$A blends through a user$-$friendly interface. In addition, the PLQ$-$Sim program can be easily updated to implement new features suggested by the users due to its modular characteristics. We are confident that the various features offered by the PLQ$-$Sim code will contribute significantly to the widespread adoption of this tool as a valuable instrument in researching more efficient organic devices based on D$/$A blends.

\section*{Declaration of Competing Interest}
\noindent The authors declare that they have no known competing financial interests or personal relationships that could have appeared to influence the work reported in this paper.

\section*{Acknowledgments}
\noindent The authors acknowledge financial support from LCNano/SisNANO 2.0 (grant 442591/2019$-$5), INCT $-$ Carbon Nanomaterials, INCT $-$ Materials Informatics, and INCT $-$ NanoVIDA. L.B. (grant E$-$26/202.091/2022 process 277806), O.M. (grant E$-$26/200.729/2023 process 285493)  and G.C. (grant E$-$26/200.627/2022 and E$-$26/210.391/2022 process 271814) are grateful for financial support from FAPERJ. The authors also acknowledge the computational support of N\'{u}cleo Avan\c{c}ado de Computa\c{c}\~{a}o de Alto Desempenho (NACAD/COPPE/UFRJ), Sistema Nacional de Processamento de Alto Desempenho (SINAPAD), Centro Nacional de Processamento de Alto Desempenho em S\~{a}o Paulo (CENAPAD$-$SP), and technical support of SMMOL$-$solutions in functionalyzed materials.

\section{Data availability}
\noindent Data will be made available on request.

 \bibliographystyle{elsarticle-num} 
 \bibliography{PLQ-Sim}

\end{document}